\documentstyle[prd,aps,epsf,eqsecnum]{revtex}
\begin{document}
\def\CC{{\Bbb C}}
\def\NN{{\Bbb N}}
\def\QQ{{\Bbb Q}}
\def\RR{{\Bbb R}}
\def\ZZ{{\Bbb Z}}
\def\cA{{\cal A}}          \def\cB{{\cal B}}          \def\cC{{\cal C}}
\def\cD{{\cal D}}          \def\cE{{\cal E}}          \def\cF{{\cal F}}
\def\cG{{\cal G}}          \def\cH{{\cal H}}          \def\cI{{\cal I}}
\def\cJ{{\cal J}}          \def\cK{{\cal K}}          \def\cL{{\cal L}} 
\def\cM{{\cal M}}          \def\cN{{\cal N}}          \def\cO{{\cal O}}
\def\cP{{\cal P}}          \def\cQ{{\cal Q}}          \def\cR{{\cal R}} 
\def\cS{{\cal S}}          \def\cT{{\cal T}}          \def\cU{{\cal U}}
\def\cV{{\cal V}}          \def\cW{{\cal W}}          \def\cX{{\cal X}}
\def\cY{{\cal Y}}          \def\cZ{{\cal Z}}

\def\arsinh{\mathop{\rm arsinh}\nolimits}
\def\arsinh{\mathop{\rm arcosh}\nolimits}
\def\arsin{\mathop{\rm arsin}\nolimits}
\newtheorem{theorem}{Theorem}
\newtheorem{prop}{Proposition}
\newtheorem{conj}{Conjecture}
\newcommand{\be}{\begin{equation}}
\newcommand{\ee}{\end{equation}}
\newcommand{\dd}{\partial}
\newcommand{\bea}{\begin{eqnarray}}
\newcommand{\eea}{\end{eqnarray}}

\title{DYNAMICAL COMPACTIFICATION, STANDARD COSMOLOGY AND THE ACCELERATING UNIVERSE}

\author{N. Mohammedi \footnote{En d\'el\'egation au CNRS.}$^{,}$\footnote{Permanent address: 
Laboratoire de Math\'ematiques et Physique Th\'eorique, 
Universit\'e Fran\c{c}ois Rabelais,
Facult\'e des Sciences et Techniques,  
Parc de Grandmont, F-37200 Tours, France.}$^{,}$\footnote{E-mail: 
nouri@celfi.phys.univ-tours.fr}}

\address{\textit{
Division of Theoretical Physics, 
Department of Mathematical Sciences,\\ 
Chadwick Building, 
The University of Liverpool, \\
Liverpool, 
L69 3BX, 
England, U.K.\\ }}

\maketitle

\vspace{5mm}

\begin{abstract}
A cosmological model based on Kaluza-Klein theory is studied. A metric, in which the scale factor
of the compact space evolves as an inverse power of the radius of the observable universe, is constructed.
The Freedmann-Robertson-Walker equations of standard four-dimensional cosmology are obtained precisely.
The pressure in our universe is an effective pressure expressed in terms of the components of the
higher dimensional energy-momentum tensor.  In particular, this effective pressure could be negative
and might therefore explain the acceleration of our present universe.
A special feature of this model is that, for a suitable choice of the parameters of the metric,
the higher dimensional gravitational coupling constant could be negative.

\end{abstract}

\smallskip
\smallskip

\centerline{February 2002}



\section{Introduction}

The idea that our four-dimensional world could have emerged, through compactification, from
a higher dimensional space-time is receiving much attention nowadays
(see \cite{kk} for a review).
However, the question of why and how does this compactification take place remains to be
answered. In string theory, considered to be a potential candidate for unifying gravity with
the other forces,  the higher dimensional manifold has to possess special properties. There,
the compactification is possible if the geometry of the manifold allows, for instance, the
existence of suitable Killing vectors.  The major difficulty is, however, to understand 
why such manifolds are favoured and whether other means for carrying out compactification
are possible.    
\par
In cosmology other types of compactifications, not necessarily relying on Killing vectors,
could be encountered. This approach is known as dynamical compactification. Here the extra dimensions
evolve in time to extremely small sizes and the universe reduces to an effective four-dimensional
one. This type of compactification was considered in the context of Kaluza-Klein theories by
Chodos and Detweiler \cite{chodos}. 
However, the use of extra dimensions in cosmology was already initiated in \cite{peter}.
The study of ref.\cite{chodos} is based on a Kasner type metric
in the vacuum and various
issues were later explored by other authors 
\cite{lot1,lot2,lot3,lot4,lot5,bailin,marciano,lot6,lot7,lot8,lot9,lot10,lot11,lot12,lot13,kolb,taylor,barr,lot14,lot15,lot16,lot17,lot18,poland,chakra,lot19,lot20,baner,sriva,kiri}.
\par  
Naturally, one would like to know how does 
this effective four-dimensional universe evolve and whether the corresponding cosmology differs
from the standard Friedmann-Robertson-Walker of ordinary space-time without extra dimensions. 
Most of the investigations in this context have dealt with finding solutions to the 
higher dimensional equations. Some of these solutions are indeed of the form found
in ordinary four-dimensional cosmology \`a la  Friedmann-Robertson-Walker. 
However, this could hardly be considered as an indication that standard cosmology
is obtainable from higher dimensions upon compactification.
\par
In this paper we will revisit the programme of dynamical compactification using a more general
metric as compared to ref.\cite{chodos}. We will also include matter without specifying its nature or origin
(see refs.\cite{lot8,free1,free2} for some derivations of the matter contribution in Kaluza-Klein cosmology).
The size of the compact space evolves as an inverse power
of the radius of our universe. It will be shown that standard Friedmann-Robertson-Walker 
cosmology is obtained.  Indeed, we are able to find an effective pressure in such a way that
the higher dimensional equations of motion  yield the usual equations of motion of ordinary
four-dimensional cosmology. There is one remaining equation which simply determines,
in terms of the radius of our universe,  the pressure along the compact dimensions. 
\par
We start, in this paper, by outlining the main points of the standard model of cosmology
in four dimensions.  In section three, the higher dimensional model is presented together with
its equations of motion. These are shown to yield the four-dimensional equations of motion stated in 
section two. {}Finally, some particular solutions are given in section four. The implications
of our model are also discussed.

\section{Standard four dimensional cosmology}

In order to compare standard cosmology with Kaluza-Klein cosmology, it is instructive to briefly
recall the Friedmann-Robertson-Walker equations.
We start from the four-dimensional action
\begin{equation}
S=\alpha\int d^4x\sqrt{-g}\left\{R -2\Lambda\right\} +S^{(4)}_{\rm{m}}\,\,\,\,,
\end{equation}
where $\alpha={1\over 16\pi G}$ in a system of units where the speed of light
$c=1$, $\Lambda$ is a cosmological constant and $S^{(4)}_{\rm{m}}$ is the
action for the matter fields.
The variation of  the above action with respect to $g^{\mu\nu}$ leads to 
\bea
2\alpha\left(R_{\mu\nu} -{1\over 2}g_{\mu\nu} R + g_{\mu\nu} \Lambda\right)
=T_{\mu\nu} \,\,\,\,,
\eea
where $T_{\mu\nu}$ is the contribution of the matter action $S^{(4)}_{\rm{m}}$ and is defined by
$\delta_gS^{(4)}_{\rm{m}}=-{1\over 2}\int d^4x\sqrt{-g} T_{\mu\nu}\delta g^{\mu\nu}$.
Standard cosmology is built on the Friedmann-Robertson-Walker metric
\be
ds^2=-dt^2 +a\left(t\right)^2\left[{dr^2\over 1-kr^2} +r^2\left(d\theta^2+\sin^2\theta d\phi^2\right)\right]
\,\,\,\,.
\ee
Here $k$ is the spatial curvature. This metric is accompanied by
the energy-momentum tensor $T^\mu_\nu$ 
\be
T^\mu_\nu={\rm diag}\left[-\varrho\left(t\right),P\left(t\right),P\left(t\right),P\left(t\right)\right]
\,\,\,\,.
\ee
The equations of motion reduce to the following two equations
\bea
{\varrho\over 2\alpha} &=& 3 H^2 +{3 k\over a^2} -{\Lambda}
\nonumber\\
{P\over 2\alpha} &=& -2{\ddot{a}\over a} - H^2 -{k\over a^2} + {\Lambda}
\,\,\,\,.
\label{4dfrw}
\eea
Here a dot denotes differentiation with respect to $t$ and $H={\dot a\over a}$ is
the Hubble parameter.  
\par
Combining these last two equations results in
\bea
6{\ddot{a}\over a}+ {1\over 2\alpha}\left(\varrho + 3P\right) -2\Lambda  =0\,\,\,.
\label{4daccel}
\eea
The combination $\left(\varrho + 3P\right)$ is a typical characteristic of four-dimensional cosmology.
This remark will be 
of use when considering the higher dimensional case.
This last equation shows, in the case of a vanishing cosmological constant\footnote{Alternatively, 
the cosmological constant 
$\Lambda$ can be absorbed by redefining the energy density and the pressure as follows:
$\varrho\longrightarrow \varrho_{\rm{tot}} -2\alpha\Lambda$ and 
$P\longrightarrow P_{\rm{tot}} +2\alpha\Lambda$.},
that the sign of the acceleration of the universe, ${\ddot a\over a}$, is given by the sign 
of the quantity $-\left(\varrho + 3P\right)$. In particular, if the universe is accelerating 
(as it is at the present time \cite{accel}) then $P<-{\varrho\over 3}$. 
It is clear that physical matter cannot be described by a negative pressure.
However, the energy-momentum tensor corresponding to an interacting scalar field can lead,
under the assumption that the scalar field depends on time only, to negative pressure.
The scalar field representation of the energy-momentum tensor in this case is known as quintessence \cite{quin}. 
\par  
Another important equation is derived from the conservation equation $\nabla_\mu T^\mu_\nu=0$.
This is a direct consequence of the Bianchi identities of the Riemann tensor in the above
equations of motion. It reads 
\be
\dot\varrho a +3\left(\varrho +P\right)\dot{a}=0
\,\,\,\,.
\ee
The physical interpretation of this last equation is better understood by rewriting it in the 
form
\be
{d\over dt}\left(a^3\varrho\right) + P {d\over dt}\left(a^3\right)=0\,\,\,.
\label{4dcons}
\ee
This means that the rate of change of total energy in a volume element of size $V=a^3$ is equal to minus
the pressure times the change of volume: $dE=-PdV$.
\par
It is this last equation which will be our guiding principle when dealing with the Kaluza-Klein case.
It allows us to define an effective pressure, $\widetilde{p}$, such that an equation
of the form  $dE=-\widetilde{p}dV$ can always be found.

\section{Kaluza-Klein cosmology}

We would like to investigate now whether the equations of the previous section 
can be derived from higher dimensions.
The starting point for this study is the action functional 
\begin{equation}
S=\beta\int d^Dx\sqrt{-{\cal{G}}}\left\{{\cal{R}} -2 \lambda\right\} +{\cal{S}}^{(D)}_{\rm{m}}\,\,\,,
\end{equation}
where $D=4+d$, ${\cal{R}}$ is the Ricci scalar corresponding to the D-dimensional metric
${\cal{G}}_{ij}$ ($i,j,\dots=0,\dots,3+d$) and $\beta$ is the higher dimensional gravitational coupling constant. 
We have also included a cosmological constant $\lambda$ and a matter action 
${\cal{S}}^{(D)}_{\rm{m}}$. 
Varying the above action with respect to ${\cal{G}}^{ij}$, we find 
\bea
{\cal{H}}_{ij} \equiv  2\beta\left({\cal{R}}_{ij} -{1\over 2}{\cal{G}}_{ij} {\cal{R}} 
+ {\cal{G}}_{ij} \lambda\right)
-{\cal{T}}_{ij}=0\,\,\,\,.
\eea
Here ${\cal{T}}_{ij}$ is the energy-momentum tensor derived from the matter action ${\cal{S}}^{(D)}_{\rm{m}}$.
\par
A cosmological setting based on Kaluza-Klein theory could be realised through a generalisation
of the homogeneous Friedmann-Robertson-Walker space-time. More specifically, the higher dimensional
metric ${\cal{G}}_{ij}$ is chosen to have the form
\be
ds^2=-dt^2 +a\left(t\right)^2\left[{dr^2\over 1-Kr^2} +r^2\left(d\theta^2+\sin^2\theta d\phi^2\right)\right]
+b\left(t\right)^2\gamma_{ab}\left(y\right) dy^ady^b
\,\,\,\,,
\ee  
where the extra coordinates are denoted by $y^a$ ($a,b,\dots=4,\dots,3+d$)
and the compact manifold is described
by the metric $\gamma_{ab}$. This spatial manifold is chosen to be maximally symmetric with
a Riemann tensor given by 
$R_{abcd}=\kappa\left(\gamma_{ac}\gamma_{bd}-\gamma_{ad}\gamma_{bc}\right)$, where $\kappa$ is a constant.
The parameter $K$ is the constant curvature of the 
spatial manifold spanned by the coordinates $\left(r, \theta , \phi\right)$.
Notice that if the scale factor $b\left(t\right)$ gets
extremely small, then the sizes of the extra dimensions shrink to zero and the space-time becomes
an effective four-dimensional one. This is the main idea behind dynamical compactification. 
\par 
The form of the energy-momentum tensor ${\cal{T}}^i_j$ 
is dictated by Einstein's equations and by the symmetries of the above metric.
In this case it is given by
\be
{\cal{T}}^i_j={\rm diag}\left[-\rho\left(t\right),p\left(t\right),p\left(t\right),p\left(t\right)
,p_d\left(t\right),\dots,p_d\left(t\right)\right]
\,\,\,\,,
\ee 
where $p_d\left(t\right)$ is the pressure along the extra dimensions.
We can now proceed to the analyses of the equations of motion. 
The component ${\cal H}^0_0$ yields the energy density
\bea
{\rho\over 2 \beta} &=&
\left\{3H^2 + {3K\over a^2} -\lambda\right\}
+{1\over 2}d\left(d-1\right){\dot{b}^2\over b^2} + 3d{\dot{a}\over a}{\dot{b}\over b}
+ {d\over 2}\left(d-1\right){\kappa\over b^2}
\nonumber\\
&\equiv& \left\{3H^2 + {3K\over a^2} -\lambda\right\}
+{\rho_0\over 2\beta}
\,\,\,, 
\label{rho}
\eea 
where $H={\dot{a}\over a}$ denotes the Hubble parameter as before.
Next we have ${\cal H}^1_1={\cal H}^2_2={\cal H}^3_3$ and we obtain 
\bea
{p\over 2\beta} &=&
\left\{-2{\ddot{a}\over a} - H^2 -{K\over a^2} + {\lambda}\right\}
-d{\ddot{b}\over b} -{1\over 2}d\left(d-1\right){\dot{b}^2\over b^2}
-2d{\dot{a}\over a}{\dot{b}\over b} -{d\over 2}\left(d-1\right){\kappa\over b^2}
\nonumber\\
&\equiv&
\left\{-2{\ddot{a}\over a} - H^2 -{K\over a^2} + {\lambda}\right\}
+{p_0\over 2\beta}
\,\,\,.
\label{p}
\eea
Finally, the components ${\cal{H}}^a_b$ lead to the single equation
\bea
{p_d\over 2 \beta}= -3{\ddot{a}\over a} -3H^2 -{3K\over a^2} +\lambda
-\left(d-1\right){\ddot{b}\over b}
-{1\over 2}\left(d-1\right)\left(d-2\right){\dot{b}^2\over b^2} 
-3\left(d-1\right){\dot{a}\over a}{\dot{b}\over b}
-{1\over 2}\left(d-1\right)\left(d-2\right){\kappa\over b^2}
\,\,\,\,.
\label{p5}
\eea
The expressions between curly brackets in the equations of $\rho$ and $p$
resemble those found on the right-hand-side of the four-dimensional
Freedmann-Robertson-Walker expressions in (\ref{4dfrw}).
Therefore, the four-dimensional energy density $\varrho$ and pressure $P$
are, respectively, identified with the quantities $\left(\rho-\rho_0\right)$ and 
$\left(p-p_0\right)$.  This is the usual and standard interpretation of the higher dimensional equations 
of motion. In this interpretation, however, $\varrho$ and $P$ contain contributions involving the scale 
factors $a$ and $b$. It seems also that the form of the scale factor $b$ does not intervene (in other words,
the resulting four-dimensional quantities do not know whether $b$ is increasing or decreasing
with time).
The natural question to be asked now is whether there are other interpretations of the higher dimensional
equations of motion. It will be shown in this note that the answer to this question is affirmative.
In particular, the four-dimensional quantities $\rho$ and $P$ are found to be given by linear combinations
of $\rho$, $p$ and $p_d$ without any explicit dependence on the scale factors $a$. This interpretation is,
however, associated with a specific choice of the scale factor $b$.
\par 
A full comparaison with the four-dimensional case, however,
requires also the determination of the equivalent
of the conservation equation (\ref{4dcons}). This could only result from the
higher dimensional conservation equation.
As a consequence of the Bianchi identities of the Riemann tensor, the equations
of motion lead to $\nabla_i {\cal{T}}^i_j=0$. Explicitly, this conservation equation reads
\be
\left\{{d\over dt}\left(a^3\rho\right) +p{d\over dt}\left(a^3\right)\right\}
+da^3{\dot{b}\over b}\left(\rho+p_d\right)=0
\,\,\,
\label{dcons}
\ee
Again, the quantity between curly brackets is of the form of the four-dimensional
conservation equation (\ref{4dcons}). In the standard interpretation of the
higher dimensional equations, this last equations takes, as expected, the form
${d\over dt}\left[a^3\left(\rho-\rho_0\right)\right] +\left(p-p_0\right)
{d\over dt}\left(a^3\right)=0$.
\par
The key point in the alternative interpretation of the higher dimensional
equations of motion resides in the last equation (\ref{dcons}). 
It is the factor in the last term  of this equation, $a^3{\dot{b}\over b}$,  which one has to handle with care. 
In the simplest of all cases, one could make this term to vanish
by choosing the scale factor $b\left(t\right)$ to be a constant
equal to $b_0$. In this situation, the conservation equation (\ref{dcons}) is equivalent to
the conservation equation of four dimensions (\ref{4dcons}). {}Furthermore,  
the expressions of $\rho$ in (\ref{rho})
and $p$ in (\ref{p}) yield the four-dimensional Friedmann-Robertson-Walker equations (\ref{4dfrw}) 
provided that we make the identifications
\bea
\alpha &=& v^{(d)}\beta\nonumber\\
\Lambda &=&\lambda -{d\over 2}\left(d-1\right){\kappa\over b_0^2}
\nonumber\\
k &=& K\,\,\,\,,
\label{firstidentif}
\eea
where $v^{(d)}$ is the finite volume of the extra $d$-dimensional spatial manifold.
The introduction of this volume is necessary in order for $\alpha$ to have the right dimensionality.
Notice that the four-dimensional cosmological constant $\Lambda$ could be adjusted to zero
by tuning the curvature $\kappa$. However, if $d=1$ then $\Lambda$ vanishes only when the
higher dimensional cosmological constant $\lambda$ is itself zero.
The components of the four-dimensional energy-momentum tensor are given by
\be
\varrho=v^{(d)}\rho\,\,\,\,\,\,\,,\,\,\,\,\,\,\,
P=v^{(d)}p \,\,\,.
\ee
The remaining equation, namely (\ref{p5}), expresses simply the pressure along the extra spatial
directions. It is given by
\bea
{p_d\over 2 \beta}= -3{\ddot{a}\over a} -3H^2  
-{3K\over a^2} +\lambda -{1\over 2}\left(d-1\right)\left(d-2\right){\kappa\over b_0^2}
\,\,\,\,.
\eea
Choosing $b\left(t\right)=b_0$ leads to simple expressions
but it is very difficult to justify. {}Furthermore, in the context of dynamical compactification,
$b$ has to be a decreasing function of time.   
\par
The second situation we would like to examin now is when the scale factor $b\left(t\right)$
evolves in time. However, the dependence on time of $b$ is chosen such that 
$a^3{\dot{b}\over b}$ is proportional to ${d\over dt}\left(a^3\right)$. This is the case if we have
\be
b={1\over a^n}\,\,\,\,.
\ee
The parameter $n$ must be positive for dynamical compactification to take place.
Therefore $b$ gets smaller as the radius of our universe $a$ becomes bigger.
\par
The fact that now $a^3{\dot{b}\over b}=-{n\over 3} {d\over dt}\left(a^3\right)$
allows one to write the conservation equation (\ref{dcons}) as
\be
{d\over dt}\left(a^3\rho\right) +\widetilde{p}{d\over dt}\left(a^3\right)=0\,\,\,,
\ee
where $\widetilde{p}$ is an effective pressure given by
\be
\widetilde{p}=p-{dn\over 3}\left(\rho +p_d\right)\,\,\,.
\label{effectivep}
\ee
The higher dimensional conservation equation is now of the same form as that of four 
dimensions (\ref{4dcons}). Let us see what becomes of the other equations.
\par
Substituting for $b$ in (\ref{rho}), the expression of the energy density becomes
\bea
{\rho\over 2\beta} &=&
{1\over 2}\left[6+dn\left(dn-n-6\right)\right]H^2 +{3K\over a^2}-\lambda 
+{1\over 2}\kappa d\left(d-1\right)a^{2n}\,\,\,\,
\label{rho1} 
\eea
while that of the pressure $p$ in (\ref{p}) leads to
\bea
{p\over 2\beta} &=& \left(dn-2\right){\ddot{a}\over a}
-{1\over 2}\left[2+dn\left(dn+n-2\right)\right]H^2-{K\over a^2} +\lambda 
-{1\over 2}\kappa d\left(d-1\right)a^{2n}\,\,\,\,.
\label{p1}
\eea
Similarly, the pressure, $p_d$ in (\ref{p5}) is found to be given by
\bea
{p_d\over 2\beta} &=& \left(dn-n-3\right){\ddot{a}\over a}
-{1\over 2}\left[6+n\left(d-1\right)\left(dn-4\right)\right]H^2 -{3K\over a^2} +\lambda
-{1\over 2}\kappa \left(d-1\right)\left(d-2\right)a^{2n}
\,\,\,.
\label{pd1}
\eea
Finally, the effective pressure, $\widetilde{p}$, is computed from these last three equations 
and is found to take the form
\bea
{\widetilde{p}\over 2\beta} &=&
-{1\over 3}\left[6+dn\left(dn-n-6\right)\right]{\ddot{a}\over a}
-{1\over 6}\left[6+dn\left(dn-n-6\right)\right]H^2 -{K\over a^2} +\lambda
-{1\over 6}\kappa d\left(2n +3\right)\left(d-1\right)a^{2n}\,\,\,.
\label{dptilde}
\eea
The terms proportional to $a^{2n}$ in the expressions of $\rho$, $p$ and $p_d$
are physically unacceptable as it means that these quantities are 
increasing with the radius of the universe. 
This can of course happen if the universe is contracting which is not the case of interest
to us here.
The solution to this  
problem is to take either $d=1$ or
$\kappa=0$. Therefore in the case of many compact dimensions, the scalar curvature of the 
compact manifold has to vanish.
\par
Let us therefore consider the case when $\kappa=0$ (if $d=1$ then $\kappa$ is automatically zero).
The terms in the expressions of $\rho$ in (\ref{rho1}) and 
$\widetilde{p}$ in (\ref{dptilde}) have, for $\kappa=0$,  the same relative factors as in the 
standard Freedmann-Robertson-Walker equations in (\ref{4dfrw}).
Indeed, the four-dimensional Freedmann-Robertson-Walker equations are recovered from equations (\ref{rho1})
and (\ref{dptilde})
provided that one makes the identifications
\bea
\alpha &=& {\beta\over 6}\left[6 +dn\left(dn-n-6\right)\right]v^{(d)}
\nonumber\\
k &=& {6K\over\left[6 +dn\left(dn-n-6\right)\right]}
\nonumber\\
\Lambda &=& {6\lambda\over\left[6 +dn\left(dn-n-6\right)\right]}\,\,\,.
\label{didentif}
\eea
The four-dimensional quantities $\varrho$ and $P$ are then identified with
\be
\varrho=v^{(d)}\rho\,\,\,\,\,\,\,,\,\,\,\,\,\,\,
P=v^{(d)}\widetilde{p} \,\,\,.
\label{didentif2}
\ee  
We are, of course, assuming here that $\left[6+dn\left(dn-n-6\right)\right]\ne 0$. 
\par
In conclusion, the three equations of motion of the higher dimensional theory (\ref{rho1}), (\ref{p1}) and (\ref{pd1}), 
can be combined in such a way that two of these, (\ref{rho1}) and (\ref{dptilde}), yield the equations
of motion of ordinary four-dimensional cosmology. The third equation, namely (\ref{pd1}), 
expresses simply the pressure along the extra dimensions in terms of the radius $a\left(t\right)$
of the observed universe. It is interesting to notice that although the sizes of the extra
dimensions evolve towards zero, the pressure along these directions does not necessarily vanish.
Furthermore, taking $\kappa=0$ and combining the expressions of $\rho$ in (\ref{rho1}) and 
$\widetilde{p}$ in (\ref{dptilde}), we obtain
\be
\left[6+dn\left(dn-n-6\right)\right]{\ddot{a}\over a} +{1\over 2\beta}\left(\rho+3\widetilde{p}\right)
-2\lambda=0
\,\,\,.
\label{daccel}
\ee
Again, this equation is of the same form as the four-dimensional equation (\ref{4daccel}). 
\par
At this stage some remarks are appropriate. We observe, from the first relation in (\ref{didentif}), 
that in order for the coupling constant $\alpha$ to be positive, 
one must have either $\left[6+dn\left(dn-n-6\right)\right]>0$ and $\beta$ positive or 
$\left[6+dn\left(dn-n-6\right)\right]<0$ and $\beta$ negative. 
The latter situation corresponds to  a `repulsive gravitational
interaction` in the higher dimensional theory! 
Furthermore, for $\left[6+dn\left(dn-n-6\right)\right]<0$, 
the signs of the four-dimensional quantities $k$ and $\Lambda$ are opposite
to those of $K$ and $\lambda$, respectively. 
Finally, for large $n$, both the four-dimensional cosmological constant $\Lambda$
and the spatial curvature $k$ tend to zero.
\par
Another observation concernes equation (\ref{daccel}). This equation, upon using (\ref{didentif}),
is written as
\bea
6{\ddot{a}\over a} + {1\over 2\alpha}\left(\rho + 3\widetilde{p}\right)v^{(d)} -2\Lambda  
=0\,\,\,
\eea
which is, as expected, of the same form as the four-dimensional equation (\ref{4daccel}).
In the case of a vanishing $\lambda$ (implying that $\Lambda=0$), 
the sign of ${\ddot a\over a}$  is given by the sign
of the quantity $-\left(\rho+3\widetilde{p}\right)$.  Therefore,
an accelerating universe is obtained when the effective pressure $\widetilde{p}$ is such that
$\widetilde{p}<-{\rho\over 3}$.
However, this is now acceptable
as $\widetilde{p}$ is only an effective pressure and all that $\widetilde{p}<-{\rho\over 3}$ means
is that $p<\left(dn-1\right){\rho\over 3} + {dn\over 3}p_d$.
Thas is, the four-dimensional pressure $\widetilde{p}$ could be negative when the
higher dimensional pressures $p$ and $p_d$ are positive.
\par
We return now to the discussion of the case when $\left[6+dn\left(dn-n-6\right)\right]=0$. 
Here we have the following two 
equations for the energy density and the effective pressure:
\bea
{\rho\over 2 \beta} &=& {3 K\over a^2} - {\lambda} \nonumber\\ 
{\widetilde{p}\over 2\beta} &=&  -{K\over a^2} + {\lambda} \,\,\,.
\label{[]=0}
\eea
These equations are of the form of the four-dimensional equations (\ref{4dfrw}) only for a radius 
$a\left(t\right)$ satisfying
\be
H^2={\omega\over a^2}+\gamma \,\,\,\,,
\ee
where $\omega$ and $\gamma$ are two constants. This conclusion is reached by comparing 
the two expressions of the energy densities in (\ref{[]=0}) and (\ref{4dfrw}). 
The general solution to this 
last differential equation, if $\gamma\ne 0$, is given by
\be
a\left(t\right)= \psi \exp\left(\sqrt{\gamma\,}\,t\right)-{\omega\over 4\psi\gamma}  
\exp\left(-\sqrt{\gamma\,}\,t\right)\,\,\,\,,
\label{n=1sol}
\ee
where $\psi$ is an integration constant.
Substituting this expression of $a\left(t\right)$ into the four-dimensional 
equations in (\ref{4dfrw}) leads to
\bea
{\varrho\over 2\alpha} &=& 3\left(\omega +k\right){1\over a^2} 
-\left({\Lambda} - 3\gamma\right)
\nonumber\\
{P\over 2\alpha} &=& -\left(\omega +k\right) {1\over a^2} +\left({\Lambda}-3\gamma\right)
\,\,\,\,.
\eea
These equations are of the form of the higher dimensional equations (\ref{[]=0}) upon the identification
\bea
\beta K v^{(d)} &=&\alpha\left(k+\omega\right) \nonumber\\
\beta\lambda v^{(d)}&=& \alpha\left(\Lambda -3\gamma\right) \,\,\,\,
\eea
together with $\varrho=v^{(d)}\rho$ and $P=v^{(d)}\widetilde{p}$.
We see that in the last equation there are only two relations for the determination of
the three parameters $\alpha$, $k$ and $\Lambda$.  
With $a\left(t\right)$ as given by (\ref{n=1sol}), the corresponding pressure along the
extra dimensions is given by
\be
{p_d\over 2 \beta}=-\left[n\left(d+2\right)\omega+3K\right]{1\over a^2} 
+ \left[{\lambda}-3\left(n+1\right)\gamma\right]
\,\,\,\,.
\ee
Here $n$ and $d$ are related by $\left[6+dn\left(dn-n-6\right)\right]=0$.
We notice, in particular, that $p_d$ can be made to vanish by taking 
$K=- n\left(d+2\right){\omega\over 3}$  
and $\lambda=3\left(n+1\right)\gamma$.
\par
The solution (\ref{n=1sol}) represents an inflationary epoch if the constant $\gamma$ is positive.
Indeed, we have ${\ddot{a}\over a}=\gamma$ and the expansion of the universe is accelerating
for a positive $\gamma$. If on the other hand $\gamma$ is negative then 
$a\left(t\right)=\sqrt{-{\omega\over\gamma}}\cos\left(\sqrt{-\gamma}t + \zeta\right)$,
where $\zeta$ is a constant and $\omega$ is assumed positive.
This oscillation in time causes the universe to expand and to collapse repeatedly.
However, this behaviour of $a\left(t\right)$ might not be suitable for
dynamical compactification.
It is also worth examining the case when $\gamma=0$.  Here the radius 
takes the form
$a\left(t\right)=\pm\sqrt{\omega}t +\chi$
for some constant $\chi$ and $\omega$ must be positive.    
\par
Finally, an equation of state relating the energy density $\rho$ and the pressure $p$ emerges
from (\ref{[]=0}). This is given by
\be
\widetilde{p}=-{1\over 3}\rho +{4\over 3}\beta\lambda\,\,\,.
\ee
This equation holds regardless of whether the universe is accelerating ($\gamma>0$)
or decelerating ($\gamma<0$). This is so because there is no relation relating
${\ddot{a}\over a}$ and $\left(\rho +3\widetilde{p}\right)$ as in (\ref{daccel}).

\section{Particular solutions in five dimensions}

One of the first solutions in the context of dynamical compactification was presented in ref.\cite{chodos}.
The model considered there is a five-dimensional one and a solution was found in the vacuum.
In our language, this solution corresponds to the particular case 
$d=1$ and $n=1$ together with setting the curvature
$K=0$ and the five-dimensional cosmological constant $\lambda=0$. In this case, the five-dimensional
expressions of $\rho$ in (\ref{rho1}) and $\widetilde{p}$ in (\ref{dptilde}) yield $\rho=\widetilde{p}=0$ regardless
of the function $a\left(t\right)$. The expression of $p_1$ ($p_1=p_d$ for $d=1$)
 in (\ref{pd1}), on the other hand, reduces to
\bea
{p_1\over 2 \beta}=-3{\ddot{a}\over a}-3 H^2 
\,\,\,\,.
\eea
When the pressure $p_1$ vanishes (since we are in the vacuum) then the radius is given by
\be
a\left(t\right)=\sqrt{\mu t+\nu}\,\,\,
\ee
where $\mu$ and $\nu$ are two integration constants. The solution of ref.\cite{chodos} is then found
by taking $\nu=0$. 
\par
We have already presented in the previous section a particular solution 
when $\left[6+dn\left(dn-n-6\right)\right]=0$.
To add to the above examples, let us investigate whether other interesting solutions can be found.
We will consider here for simplicity the five-dimensional model only\footnote{We are assuming 
that $n\ne 1$ as the pair of values $d=1$ and $n=1$ is a solution to $\left[6+dn\left(dn-n-6\right)\right]=0$
and this case has been discussed before.}. 
We start by examining the case for which the pressure along the fifth dimension vanishes, $p_1=0$. This is 
dictated by the fact that the fifth dimension in our model tends to zero with time. It is therefore
natural to require that $p_1=0$. The equation $p_1=0$ takes the form
\be
{p_1\over 2\beta}=-3{\ddot{a}\over a} -3H^2 -{3K\over a^2}+\lambda=0\,\,\,\,.
\ee
This equation can be simplified by making the change of variables
\be
a\left(t\right)=\sqrt{h\left(t\right)}\,\,\,.
\ee
Our differential equation, $p_1=0$,  reduces then to
\be
3 \ddot{h} - 2\lambda h  +6K =0\,\,\,.
\label{p5=0}
\ee
The solution to this equation, if $\lambda\neq 0$, is given by
\be
h\left(t\right)={3K\over \lambda}+
B\exp\left(\sqrt{{2\over 3}\lambda\,}\,t\right)
+C\exp\left(-\sqrt{{2\over 3}\lambda\,}\,t\right)\,\,\,,
\label{h}
\ee
where $B$ and $C$ are two integration constants.
The corresponding expression for the energy density in (\ref{rho1}) is found to be
\be
{\rho\over 2 \beta}=-{1\over 2\lambda}\left(n-1\right)\left(9K^2-4BC\lambda^2\right){1\over h^2}
+{3nK\over h}-\left(n+1\right){\lambda\over 2}
\,\,\,.
\ee
Of course the different parameters in this last expression are such that $\rho$ is positive. 
Similarly, the expression of the effective pressure in (\ref{dptilde}) yields
\be
{\widetilde{p}\over 2\beta}= -{1\over 6\lambda}\left(n-1\right)\left(9K^2-4BC\lambda^2\right){1\over h^2}
-{nK\over h}+\left(n+1\right){\lambda\over 2}
\,\,\,\,.
\ee
By expressing $h$ in terms of $\rho$ and replacing it in the expression
of $\widetilde{p}$, we get the following equation of state
\be
\widetilde{p} = {1\over 3}\rho -4\beta\left[{nK\over h}-\left(n+1\right){\lambda\over 3}\right]\,\,\,,
\ee
where $h$ is given in terms of $\rho$ by the expression
\be
h={\beta\over \left[\rho +\left(n+1\right)\beta\lambda\right]}\left\{
3nK \pm \sqrt{9n^2K^2 -{1\over \beta\lambda}
\left(n-1\right)\left(9K^2-4BC\lambda^2\right)
\left[\rho +\left(n+1\right)\beta\lambda\right]}\right\}
\ee
Notice that if $K=0$ then this equation of state is that of a radiation
dominated universe in the presence of a cosmological constant equal to $\beta\left(n+1\right)\lambda$.
Finally, the acceleration of the universe is given by
\be
{\ddot{a}\over a}= {1\over 6}\left[\lambda -{1\over \lambda}\left(9K^2-4BC\lambda^2\right)
{1\over h^2}\right]
\,\,\,
\ee
with $h\left(t\right)$ as given in (\ref{h}). Hence, one could not in general determine
the sign of this acceleration. An accelerating universe, for example,  is obtained
for $\left(9K^2-4BC\lambda^2\right)<0$ and a positive $\lambda$.
In this case the effective pressure $\widetilde{p}$ is negative which simply means that
$p<\left(n-1\right){\rho\over 3}$. We should mention here that if $n<1$ then $\beta$
is positive while $n>1$ forces $\beta$ to be negative. {}Finally, 
a negative cosmological constant $\lambda$ leads to an 
oscillatory regime for the radius $a=\sqrt{h}$. 
\par
We have assumed so far that $\lambda\ne 0$. Let us now discuss the case when $\lambda=0$. With this value of 
$\lambda$, the solution to equation (\ref{p5=0}) is given by
\be
h\left(t\right)=-Kt^2 +zt+w\,\,\,
\label{secondh}
\ee
for some two integration constants $z$ and $w$. The corresponding expressions for $\rho$ and $\widetilde{p}$
are given by
\bea
{\rho\over 2\beta} &=& -{3\over 4}\left(n-1\right)\left(4K\omega+z^2\right){1\over h^2} +{3nK\over h}   
\nonumber\\
{\widetilde{p}\over 2\beta} &=&
-{1\over 4}\left(n-1\right)\left(4K\omega+z^2\right){1\over h^2} -{nK\over h}
\,\,\,\,,
\eea
where $h\left(t\right)$ is as given by (\ref{secondh}). The equation of state relating $\rho$ and $\widetilde{p}$
is of the form
\be
\widetilde{p}={1\over 3}\rho -{4\beta n K\over h}\,\,\,.
\ee
The acceleration of the universe is found to be given by
\be
{\ddot{a}\over a}= -{1\over 4}\left(4K\omega+z^2\right){1\over h^2}
\,\,\,.
\ee
The universe is, in particular, in a period of deceleration 
and is radiation dominated if $K=0$.
\par
The other particular case we would like to study corresponds to the situation when $p_1=-\rho$. This choice 
is interesting as it leads to $\widetilde{p}=p$. 
The general solution to the differential equation $p_1=-\rho$,
which reads ${\ddot{a}\over a} +nH^2=0$, is 
\be
a\left(t\right)=\left[\left(n+1\right)\left(q t +u\right)\right]^{{1\over n+1}}\,\,\,,
\label{a}
\ee
where $q$ and $u$ are two integration constants. This expansion is, however, a slow one.
The corresponding expressions for the energy density $\rho$ and the effective pressure $\widetilde{p}$ are
\bea
{\rho\over 2\beta} &=& -{3 q^2\left(n-1\right)\over a^{2\left(n+1\right)}}
+{3K \over a^2} - \lambda
\nonumber\\
{\widetilde{p}\over 2\beta} &=&
-{q^2\left(n-1\right)\left(2n-1\right)\over a^{2\left(n+1\right)}}
-{K\over a^2} + \lambda
\,\,\,
\eea
with $a\left(t\right)$ as given in (\ref{a}).
We notice that the energy density $\rho$ and the effective pressure $\widetilde{p}$ are, when $n\neq {1\over 2}$,
related by
\be
\widetilde{p}={1\over 3}\left(2n-1\right)\rho -{2\beta\over 3}\left[
{6n K\over a^2} -2\left(n+1\right)\lambda\right]
\,\,\,\,.
\ee
On the other hand when $n={1\over 2}$ then we have
\be
\widetilde{p}=-{\rho\over 3} + {\beta}\left[{q^2\over a^3}+{4\over 3}\lambda\right]
\,\,\,.
\ee
The acceleration of the universe is given, for any value of $n$, by the expression
\be
{\ddot{a}\over a}=-{nq^2\over a^{2\left(n+1\right)}}
\,\,\,\,.
\ee
The universe, for this solution, is slowly expanding and going through a deceleration
phase.

\section{Discussion}

We have found, in the context of Kaluza-Klein cosmology, a higher dimensional metric 
which dynamically evolves towards an effective four-dimensional one. 
This metric depends on two scale factors: one is interpreted as the radius
of our universe while the other tends to zero with time.
In addition, the compact space has to be flat ($\kappa=0$).
The remarkable feature of the model constructed here is that it gives exactly the
four-dimensional Friedmann-Robertson-Walker equations of standard cosmology.
The component of the four-dimensional energy-momentum tensor are expressed 
as linear combinations of their higher dimensional counterparts. 
In particular, the pressure in our universe $\widetilde{p}=p-{dn\over 3}\left(\rho+p_d\right)$
could be negative when the higher dimensional quanties $\rho$, $p$ and $p_d$ are all
positive.  This fact could provide an explanation for the negative pressure needed to account for
the present acceleration of our universe. This would be an alternative to the quintessence scalar
field.  
\par
The parameters of the four-dimensional world $\left(\alpha,k,\Lambda\right)$ are given in terms 
of the parameters of the higher dimensional theory $\left(\beta,K,\lambda,n,d\right)$. 
We noticed that peculiar properties appear when the parameter $n$ is in the range
\be
{1\over d-1}\left[3-\sqrt{3+{6\over d}}\right]<n<
{1\over d-1}\left[3+\sqrt{3+{6\over d}}\right] 
\ee
and $d\ne 1$. {}First of all the higher dimensional gravitational constant $\beta$ has to be 
negative for the four-dimensional one, $\alpha$,  to be positive. In the same range of $n$,
the curvature $K$ and the cosmological constant $\lambda$ are of opposite signs to the 
four-dimensional quantities $k$ and $\Lambda$.   
In order to have a very rapid dynamical dimensional reduction, large values of $n$ are
needed. In this case (when $n\longrightarrow \infty$), the resulting four-dimensional
universe is spatially flat ($k\longrightarrow 0$) with a vanishing cosmological constant
($\Lambda\longrightarrow 0$).
\par
The five-dimensional case ($d=1$) deserves a special mention. Here, the five-dimensional
coupling constant $\beta$ is negative for $n>1$ (that is for a rapidely decreasing $b$).
Similarly, for $n>1$, the constants $K$ and $\lambda$ have the opposite signs of $k$ and
$\Lambda$.

\smallskip
\smallskip
\smallskip
\bigskip

\noindent $\underline{\hbox{\bf Acknowledgments}}$: I would like to thank the members of the 
Division of Theoretical Physics at the University of Liverpool for their hospitality.  
This research is supported by the CNRS.

\end{document}